\documentclass[preprint,epsfig,aps,showpacs]{revtex4}
\usepackage{epsfig}
\begin{document}
\newcommand{\Be}{\begin{equation}}
\newcommand{\Ee}{\end{equation}}
\newcommand{\Bea}{\begin{eqnarray}}
\newcommand{\Eea}{\end{eqnarray}}

\setcounter{page}{0}

\title{3+2 neutrinos in a see-saw variation}
\author{G. J. Stephenson, Jr.}\email{ GJS@baryon.phys.unm.edu}
\affiliation{Department of Physics and Astronomy, 
University of New Mexico, Albuquerque, NM 87131}

\author{T. Goldman}\email{ tgoldman@lanl.gov}
\affiliation{Theoretical Division, MS-B283, 
Los Alamos National Laboratory, Los Alamos, NM 87545}
\author{B. H. J. McKellar}\email
{ b.mckellar@physics.unimelb.edu.au} 
\author{M. Garbutt}\email{ mgarbutt@treasury.gov.au}
\affiliation{University of Melbourne 
Parkville, Victoria 3052, Australia}

\begin{flushright}
\vspace{-1.5in}
{LA-UR-03-5159}\\
\vspace{-0.1in}
{hep-ph/0307245}\\
\vspace*{0.5in}
\end{flushright}

\begin{abstract}

If the sterile neutrino mass matrix in an otherwise 
conventional see-saw model has a rank less than the 
number of flavors, it is possible to produce pseudo-Dirac 
neutrinos. For the rank 1 case, 3+2 scenarios devolve 
naturally as we show by example. Additionally, we 
find that the lower rank see-saw suppresses some mass 
differences, so that small mass differences do not 
require that the individual masses of each neutrino 
must also be small. 
\end{abstract}

\maketitle

\section{Introduction}

Conventional wisdom holds that neutrinos ought to be Majorana 
particles with very small masses, due to the action of a ``see-saw''
mechanism~\cite{see-saw}, which is built on the concept of 
quark-lepton symmetry.  Alternatively, there have been recent 
theoretical suggestions~\cite{GSMcK,KC} that neutrinos may 
well be Majorana particles occuring in nearly degenerate pairs, 
the so-called pseudo-Dirac neutrinos.  Recent results from 
Kamiokande~\cite{superK} on atmospheric neutrinos, from 
Sudbury~\cite{SNO} on solar neutrinos, and from KamLand~\cite{KamL}
on long baseline reactor neutrinos, appear to require
oscillations between nearly maximally mixed mass eigenstates.
Each of these analyses, however, argues that this mixing cannot 
be dominantly to sterile states such as are found in pseudo-Dirac 
pairs. On the other hand, the concatenation of the data from 
these experiments with that from LSND~\cite{LSND} and other 
short baseline data does not fit into a theoretical structure 
which only includes mixing among three active Majorana neutrinos. 
Many have therefore been motivated to consider the effects of 
additional (sterile, Majorana) neutrino states, the existence 
of which is accepted in the conventional ``see-saw'' extension 
of the standard model (SM), although there, the actual states 
are precluded, due to assumed large masses, from appearing 
directly in experiments. 

It should be noted that there is no accepted principle 
that specifies the flavor space structure of the mass 
matrix assumed for the sterile sector. Some early 
discussions~\cite{wolfm2} implicitly assume that a 
mass term in the sterile sector should be proportional 
to the unit matrix.  This has the pleasant prospect, 
in terms of the initial argument for the see-saw, that 
all neutrino flavors have small masses on the scale of 
other fermions.  However, since there is no obvious 
requirement that Dirac masses in the neutral lepton 
sector are the same as Dirac masses in any other 
fermionic sector, this result is not compelling.  
Indeed, Goldhaber has argued for a view of family 
structure and self-energy based masses that naturally 
produces small neutrino masses~\cite{Maurice}. We 
discuss here a more conventional possibility which 
arises from a minimal modification of the standard see-saw, 
namely that the rank of the mass matrix for the sterile 
sector is less than the number of flavors. Note that 
this does not conflict with quark-lepton symmetry which 
applies only to the number and character of states. 

In this paper, which is an extension of reference~\cite{hep}, 
we shall concentrate on the case of a rank $1$ sterile matrix, 
relegating the rank $2$ case to some remarks at the end. (The 
analysis of short baseline data by Sorel, Conrad and 
Shaevitz~\cite{Sorel} suggests that the rank $2$ case may not 
actually occur in Nature.) We also consider rank $1$ to be the 
more natural case because whatever spontaneous symmetry breaking 
produces mass in the $3$-dimensional sterile flavor space 
necessarily defines a specific direction. Before including the 
effects of the sterile mass, we assume three non-degenerate 
Dirac neutrinos, (although this is not essential,) which are 
each constructed from one Weyl spinor which is active under the 
$SU(2)_W$ of the SM and one Weyl spinor which is sterile under 
that interaction. (Being neutrinos, both Weyl fields have no 
interactions under the $SU(3)_C$ or the $U(1)$ of the SM.) 
There is then an MNS~\cite{MNS} matrix which relates these 
Dirac mass eigenstates to the flavor eigenstates in the usual 
manner.  Note, however, that these matrix elements are not 
those extracted directly from experiment, as the mass matrix 
in the sterile sector induces additional mixing.

We next use the Dirac mass eigenstates to define bases in both 
the $3$-dimensional active flavor space and the $3$-dimensional 
sterile flavor space~\cite{fn1}. Following the spirit of the original 
see-saw, we exclude any initial Majorana mass term in the active 
space. If the Majorana mass matrix in the sterile space were to 
vanish also, the three flavors of Dirac neutrinos would be a 
mixture of (Dirac) mass eigenstates in a structure entirely 
parallel to that of the quarks. 

A rank $1$ sterile mass matrix may be represented as a vector 
of length $M$ oriented in some direction in the $3$-dimensional 
sterile space. If that vector lies along one of the axes, then 
the Dirac neutrino that would have been formed from it and its 
active neutrino partner will partake of the usual see-saw 
structure (one nearly sterile Majorana neutrino with mass 
approximately $M$ and one nearly purely active neutrino with 
mass approximately $m_D^2/M$) and the other two mass eigenstates 
will remain Dirac neutrinos. If that vector lies in a plane 
perpendicular to one axis, the eigenstate associated with that 
axis will remain a pure Dirac neutrino, and the other two will 
form one pseudo-Dirac pair and a pair displaying the usual 
see-saw structure. Both of these pairs will be mixtures of the 
$4$ Weyl fields associated with the two mixing Dirac neutrinos. 
In general, the structure will be $2$ pseudo-Dirac pairs and one 
see-saw pair, all mixed.

As we implied above, the very large mixing required by the 
atmospheric neutrino measurements could have been taken to be 
evidence for a scheme involving pseudo-Dirac neutrinos. (This, 
after all, follows Pontecorvo's initial suggestion~\cite{BP}.) 
However, pure mixing into the sterile sector is now strongly 
disfavored~\cite{nsm}.  It is evident from the discusion above 
that there is a region of parameter space (directions of the 
vector) in which the two pseudo-Dirac pairs are very nearly
degenerate, giving rise to the possibility of strong mixing in 
the active sector coupled with strong mixing into the sterile 
sector.  We shall explore this point here.

The organization of the remainder of the paper is as follows: 
In the next section we present the mass matrix, discuss the 
parameterization of the sterile mass matrix and various 
limiting cases.  We show the spectrum for a general case.  
In the section following that, we specialize to the case 
where the vector representing the sterile mass entry lies 
in a plane perpendicular to one of the axes.  In this case 
we can carry out an analytical expansion in the small 
parameter $<m_D>/M$.  In the fourth section we apply those 
results to the case where the plane in question is perpendicular 
to the axis for the middle value Dirac mass eigenstate, raising 
the possibility of near degeneracy  between pseudo-Dirac pairs. 
Moving away from that plane produces large  mixing amongst the 
members of those pseudo-Dirac pairs.  Finally, we remark on the 
structures expected for a rank $2$ sterile matrix and then 
reiterate our conclusions.

\section{General mass matrix}

The flavor basis for the active neutrinos and the pairing to 
sterile components defined by the (generally not diagonal) 
Dirac mass matrix could be used to specify the basis for the 
sterile neutrino mass matrix, $M_S$. Instead we take the basis 
in the sterile subspace to allow the convention described 
below. This implies a corresponding transformation of the 
Dirac mass matrix, which is irrelevant at present since the 
entries in that matrix are totally unknown. 

We now define our convention for the choice of axes in the 
sterile subspace. Denote the nonzero mass eigenvalue of 
the rank $1$ by $M$ and choose its eigenvector initially 
in the third direction. Then rotate this vector, first by 
an angle of $\theta$ in the $1-3$ plane and then by $\phi$ 
in the $1-2$ plane. The rotation is chosen so that the 
Dirac mass matrix which couples the active and sterile 
neutrinos becomes diagonal, i.e., the basis is defined by 
Dirac eigenstates. This produces a $3 \times 3$ mass matrix 
in the sterile sector denoted by
\Be
M_S = M \left[ 
\begin{array}{ccc}
\cos^2 \phi \sin^2 \theta  &  \cos \phi \sin \phi \sin^2 \theta  &
\cos \phi \sin \theta \cos \theta  \\  \cos \phi \sin \phi \sin^2 \theta & 
\sin^2 \phi \sin^2 \theta  &  \sin \phi \sin \theta \cos \theta  \\ 
\cos \phi \sin \theta \cos \theta & \sin \phi \sin \theta \cos \theta & 
\cos^2 \theta   \end{array}  \right]. 
\Ee

In this representation, the Dirac mass matrix is diagonal 
by construction
\Be
m_D = \left[ \begin{array}{ccc}
m_1   &  0  &  0  \\  0  &  m_2  &  0  \\  0  &  0  &  m_3
\end{array} \right].  
\Ee

Note that there are special cases.  For $\theta = 0$ and any 
value for $\phi$,
\Be
M_S = \left[ \begin{array}{ccc}
0 & 0 & 0 \\ 0 & 0 & 0 \\ 0 & 0 & M \end{array} \right]  .
\Ee
For $\theta = \pi / 2$ and $\phi = 0$, 
\Be
M_S = \left[ \begin{array}{ccc} 
M & 0 & 0 \\ 0 & 0 & 0 \\ 0 & 0 & 0 \end{array} \right], 
\Ee
and, for $\theta = \pi / 2$ and $\phi = \pi / 2$, 
\Be
M_S = \left[ \begin{array}{ccc} 0 & 0 & 0 \\ 0 & M & 0 \\ 0 & 0 & 0
\end{array} \right]. 
\Ee
These are equivalent under interchanges of the definition 
of the third axis. 

The $6 \times 6$ submatrix~\cite{fn2} of the full $12 \times 
12$ is, in block form, 
\Be
{\cal M} = \left[ \begin{array}{cc} 0 & m_D \\ m_D & 
M_S \end{array} \right].   
\Ee
Since we are ignoring CP violation here, no adjoints or 
complex conjugations of the mass matrices appear. 

Note that, in the chiral representation, the full $12 \times 12$ matrix is
\Be
\left[ \begin{array}{cc} 0 & {\cal M} \\ 
{\cal M} & 0 \end{array} \right]. \nonumber
\Ee
Thus the full set of eigenvalues will be $\pm$ the eigenvalues of
${\cal M}$.  Where it matters for some analysis we keep track of 
the signs of the eigenvalues, however for most results we present 
positive mass eigenvalues.

After some algebra, we obtain the secular equation 
\Bea
0 & = & \lambda^6 -M \lambda^5 -(m_1^2 + m_2^2 + m_3^2) \lambda^4 \nonumber \\
& & + M [m_3^2 \sin^2 \theta + m_2^2 (\sin^2 \theta \cos^2 \phi + \cos^2 \theta)] 
\lambda^3 \nonumber \\ & & + (m_1^2 m_2^2 + m_2^2 m_3^2 + m_3^2 m_1^2) 
\lambda^2 \\
& & - M (m_1^2 m_2^2 \cos^2 \theta + m_2^2 m_3^2 \cos^2 \phi \sin^2 \theta 
\nonumber \\
  &  & + m_3^2 m_1^2 \sin^2 \phi \sin^2 \theta) \lambda \nonumber \\
  &  & - m_1^2 m_2^2 m_3^2. \nonumber
\Eea  

This may be rewritten as 
\Bea
 0 & = & (\lambda^2 - m_1^2) (\lambda^2 - m_2^2) 
(\lambda^2 - m_3^2) \nonumber \\ &   &    
- \lambda M \left( \lambda^4 -\left[ m_3^2 \sin^2 \theta 
+ m_2^2 (\sin^2 \theta \cos^2 \phi + \cos^2 \theta ) \right. \right. \nonumber \\
&  & \left. \left. + m_1^2 ( \sin^2 \theta \sin^2 \phi + \cos^2 \theta) 
\right] \lambda^2 \right. \\
&  & \left. +m_1^2 m_2^2 \cos^2 \theta + m_2^2 m_3^2 \sin^2 \theta 
cos^2 \phi  \right. \nonumber \\
&  & \left. + m_3^2 m_1^2 \sin^2 \theta \sin^2 \phi \right). \nonumber
\Eea

The special cases follow directly.  For $\theta = 0$, we find
\Be
(\lambda^2 - m_1^2) (\lambda^2 - m_2^2)
(\lambda^2 - M \lambda - m_3^2) = 0, 
\Ee
for $\theta = \pi / 2$ and $ \phi = 0$
\Be
(\lambda^2 - m_2^2) (\lambda^2 - m_3^2)
(\lambda^2 - M \lambda - m_1^2) = 0, 
\Ee
and for $\theta = \pi / 2$ and $\phi = \pi / 2$
\Be
(\lambda^2 - m_3^2) (\lambda^2 - m_1^2)
(\lambda^2 - M \lambda - m_2^2) = 0. 
\Ee

If $m_1^2 = m_2^2 = m_3^2 = m^2$, then we find
\Be
(\lambda^2 - m^2)^2 (\lambda^2 - M \lambda - m^2) = 0. 
\Ee

Due to the wide range of possibilities inherent in the system, 
it is useful to examine specific numerical examples.  For the
current exercise, we have picked the following parameters:

\Bea
M & = & 1000 \nonumber \\ m_1 & = & 1 \nonumber \\ 
m_2 & = & 2 \nonumber \\ m_3 & = & 3. \nonumber
\Eea

For this choice, the eigenvalues have a definite pattern 
for all values of $\theta$ and $\phi$.  There are two very 
close pairs, with mass eigenvalues between $1$ and $3$.  
There is one very small eigenvalue, of order $10^{-3}$ 
reflecting the ratio of $m$ to $M$, and one large eigenvalue 
of order $10^{3}$ (i.e., of order $M$).  Treating the last 
two as a pair despite their disparity in mass allows us to 
present results as three tables, one for each pair, for sets 
of angles $\theta , \phi = \pi / 8, \pi / 4, 3 \pi / 8 $.

First, for the lower mass close pair, we have
\Be
\begin{array}{lccc}
\theta \backslash \phi & \pi /8  &  \pi /4  & 3 \pi / 8  \\
             &         &          &            \\
\pi / 8      & 1.398125 & 1.230175 & 1.068477   \\
             & 1.394934 & 1.228025 & 1.067688    \\
             &          &          &            \\
\pi / 4      & 1.809478 & 1.478863 & 1.151936   \\
             & 1.808183 & 1.477134 & 1.150941   \\
             &          &          &             \\
3 \pi / 8    & 1.877166 & 1.562977 & 1.18999    \\
             & 1.876742 & 1.561911 & 1.189146  \end{array}
\Ee

Then, for the next mass pair with close eigenvalues, 
we find
\Be
\begin{array}{lccc}
\theta \backslash \phi & \pi / 8  &  \pi / 4 & 3 \pi / 8   \\
             &          &          &             \\
\pi / 8      & 2.038992 & 2.107688 & 2.158044    \\
             & 2.038729 & 2.107156 & 2.157407    \\
             &          &          &              \\
\pi / 4      & 2.347974 & 2.46348  & 2.529128    \\
             & 2.346047 & 2.462176 & 2.52809     \\
             &          &          &              \\
3 \pi / 8    & 2.816525 & 2.847539 & 2.868607    \\
             & 2.815691 & 2.846972 & 2.868186   
\end{array} 
\Ee

Finally, even though it does not directly impact the 
argument, we display the remaining pair in order to 
present a complete set.  
\Be
\begin{array}{lccc}
\theta \backslash \phi & \pi / 8  &  \pi / 4 & 3 \pi / 8   \\
             &          &          &             \\
\pi / 8      & 1000.008 & 1000.008 & 1000.008    \\
             & 0.00444  & 0.005366 & 0.006778    \\
             &          &          &              \\
\pi / 4      & 1000.005 & 1000.006 & 1000.006    \\
             & 0.001997 & 0.002717 & 0.004248    \\
             &          &          &              \\
3 \pi / 8    & 1000.003 & 1000.003 & 1000.004    \\
             & 0.001289 & 0.001819 & 0.003092   
\end{array} 
\Ee

\section{Two flavor subspace}

In the next section we shall discuss the case where two 
of the pseudo-Dirac pairs are nearly degenerate and follow 
the mixing patterns as we move away from that region of 
parameter space.  To facilitate that discussion, and to 
explore a system where analytic approximations are available, 
we find it useful to examine the limit where one Dirac mass 
eigenstate remains uncoupled from all of the other states.  
Anticipating the following section, we decouple $m_2$.  This 
is equivalent to examining a two flavor system in which the 
Dirac mass eigenvalues are $m_1$ and $m_3$ and the vector 
describing the sterile mass is described by $\phi = 0$.

It is convenient to define some new symbols:
\Bea
m_0^2  & = & m_1^2 c^2  + m_3^2 s^2  \\
a & = &\frac{ \left(m_1^2 - m_3^2\right) s c }{m_0\sqrt{2}} \\
b & = & \frac{m_1m_3}{m_0}
\Eea
where $c = \cos\theta$, $s = \sin\theta$. Note the additional 
$1/\sqrt{2}$ factor in the defintion of~$a$.

Unitary transformation of the mass matrix
\Be
{\cal{M}}_{init} =
\left(\begin{array}{cccc}
0 & 0 & m_{1} & 0 \\
0 & 0 & 0 & m_{3} \\
m_{1} & 0 & Ms^{2} & Mcs \\
0 & m_{3} & Mcs & Mc^{2}
\end{array}
\right)
\Ee
into the form
\Be
{\cal{M}}_{fin} =
\left(\begin{array}{cccc}
m_0 & 0 & 0 & a \\
0 & -m_0 & 0 & -a \\
0 & 0 & 0 & b \\
a & -a & b & M
\end{array}
\right)
\Ee
makes it apparent that, to lowest order, the three small 
eigenvalues are $\pm m_0, 0$.  Note the minus sign on 
the $a$ in the (2,4) and (4,2) positions. 

The matrix, $\Omega$, which effects the transformation, 
${\cal{M}}_{fin} = \Omega^{\dag} {\cal{M}}_{init} \Omega$, 
to such a form is
\Be
\Omega =
m_{0}^{-1}\left(\begin{array}{cccc}
m_{1}c/\sqrt{2} & -m_{1}c/\sqrt{2} & m_{3}s & 0 \\
-m_{3}s/\sqrt{2} & m_{3}s/\sqrt{2} & m_{1}c & 0 \\
m_{0}c/\sqrt{2} & m_{0}c/\sqrt{2} & 0 & m_{0}s \\
-m_{0}s/\sqrt{2} & - m_{0}s/\sqrt{2} & 0 & m_{0}c
\end{array} \right)
\Ee

This suggests that writing the characteristic equation as
\Be
\mu \left(m_{0}^{2} - \mu^{2}\right) (M - \mu) =
  2 \mu^{2} a^{2} - \left(m_{0}^{2} - \mu^{2}\right) b^{2}
\Ee
which is convenient for iterative solution in a series in 
$M^{-1}$.  The usual form of the secular equation obtained 
directly from $\left| {\cal{M}}_{init} - \mu \right| = 0$, 
is
\Be
\mu^{4} - \mu^{3}M - \mu^{2}\left(m_{1}^{2} + m_{3}^{2}\right) 
+ \mu m_{0}^{2} M + m_{1}^{2} m_{3}^{2} = 0,
\Ee
which is, of course, the same equation.

The solution to order $M^{-2}$ is 
\Bea
\mu_1 & = & m_0 - \frac{a^{2}}{M} -
\frac{a^{2}}{m_{0}M^{2}}\left(m_{0}^{2} - \frac{a^{2}}{2} -
b^{2}\right)\\
\mu_2 & = & -m_0 - \frac{a^{2}}{M} +
\frac{a^{2}}{m_{0}M^{2}}\left(m_{0}^{2} - \frac{a^{2}}{2} -
b^{2}\right) \\
\mu_3 & = & - \frac{b^2}{M} \\
\mu_4 & = & M + \frac{b^2}{M} + 2 \frac{a^2}{M}
\Eea

Notice that the eigenvalues sum to $M$ as they must; however, 
the $\pm m_{0}$ leading order eigenvalues are shifted in the 
same direction at $O(M^{-1})$ and in opposite directions at 
$O(M^{-2})$.  The solutions near $\pm m_{0}$ now specialize 
to the simpler case discussed in Ref.\cite{GSMcK}. The terms 
at $O(M^{-2})$ effectively only change the value of $m_{0}$. 
Note that $\mu_{3}$ and $\mu_{4}$, do not acquire $O(M^{-2})$ 
corrections; their next correction is at the next higher order.

\section{Two nearly degenerate psudo-Dirac pairs}

Applying the techniques of the last section, we find the angle 
$\theta$ such that $m_2$ and the eigenvalue for the pseudo-Dirac 
pair above, $m_0$, are approximately degenerate.  We then vary  
$\phi$ away from $0$ and display the eigenfunctions.  To 
illustrate the general nature of the result, we have changed 
the Dirac masses from the even spacing used above.

In the Table, the Dirac masses are taken to be $m_1 =  1$, $m_2 
=  1.1$, and  $m_3 =  3$.  This effectively means that $m_1$ is 
taken to set the pseudo-Dirac mass scale.  In order to display 
the structure of the spectrum, we have chosen $M = 1000$, rather 
than a larger value, expected to be more realistic, but which 
would suppress the difference scale between the pairs.  The 
angles are given in degrees.

The Table represents only a small part of the available parameter 
space and is chosen to display some interesting possible features. 
First, $\theta$ has been chosen so that, at $\phi = 0$, the Dirac 
pair at $m_2$ is bracketed by the pseudo-Dirac pair.  Such a value 
of $\theta$ exists for any pattern of the Dirac masses.  Then, for 
small values of $\phi$, there are always two nearly degenerate 
pseudo-Dirac pairs.

Note that, for $\phi = 0$, there is no mixing between the 
field labelled by $2$ and the remaining fields, while for 
the next entry at $\phi = 2.25$ degrees there is considerable 
mixing.  That mixing increases with $\phi$ as the difference 
bewteen the eigenvalues increases.  The pattern described by 
the centroids of the pseudo-Dirac pairs is fixed by the angles 
$\theta$ and $\phi$.  If $M$ is increased, that pattern hardly  
changes.  The primary effect of increasing $M$, consistent 
with the analysis in the previous section, is to decrease the 
separation of the two members of each pseudo-Dirac pair while 
producing the usual see-saw behavior for the remaining pair. 
Thus, tiny differences in mass between masses that are not 
especially small themselves, are, in the usual sense of the 
term, natural, in this approach. 

The implication for oscillation phenomena is clear.  A given 
weak interaction produces an active flavor eigenstate which 
is some linear combination of the three active components 
listed in the Table.  That then translates into a linear 
combination of the six mass eigenstates.  From the Table, it 
is clear that the involvement of the heavy Majorana see-saw 
state is minimal, so the system effectively consists of the 
light Majorana see-saw state and the four Majorana states 
arising from the two pseudo-Dirac pairs. These five states 
include all three active neutrinos, generating a natural 
3+2 scenario. 

Since these five mass eigenstates have both active and 
sterile components, the subsequent time evolution will 
involve both flavor changing oscillations and oscillation 
into (and back out of) the sterile sector. This can lead 
to very complex oscillation patterns, as there are $10$ 
mass differences, $4$ of which are independent. For example, 
a short baseline neutrino oscillation experiment may find 
that data are better fit by a highly non-sinusoidal function. 
An example of an appearance probability for the second set 
of angle parameters in the Table is shown in the Figure. 

Finally, inspection of the column labelled ``1active'' for 
$\phi = 2.25$ or $\phi = 4.5$, for example, shows that 
the presence of a rank $1$ sterile mass matrix can 
seriously change any mixing pattern of the MNS type~\cite{MNS}, 
from that which would have obtained with purely Dirac neutrinos.

\section{Rank $2$}

We have not discussed the case of rank $2$ matrices explicitly,
although the pattern is obvious.  In such a case, there would 
be two see-saw pairs and one pseudo-Dirac pair, leading to 
three active and one sterile light neutrino.  While this pattern 
has been analyzed in the literature, we do not find any compelling 
pattern for it in the sterile sector.  Furthermore, the current 
consistency of all neutrino oscillation data can be accomodated 
much more easily (and perhaps only, as indicated by Ref.\cite{Sorel},) 
in the rank $1$ case discussed in this paper. Therefore we leave 
the discussion of rank $2$ to another time.

\section{Conclusions}

We have considered here the effects on neutrinos in the SM of the 
recurrently successful  and conventional constraint of quark-lepton 
symmetry: The existence of six independent Weyl spinor fields of 
neutrinos, three corresponding to active and three corresponding 
to sterile neutrinos. In the now venerable see-saw approach, the 
latter three effectively disappear from the excitation spectrum, 
leaving small Majorana masses for the active states as a residuum. 
We have generalized this by considering the effect on the system 
of a rank less than three character of the $3 \times 3$ mass 
matrix in the sterile sector and studied the rank $1$ case, in 
particular. 

In the rank $1$ case that we have focused on, we find that the 
neutrino fields naturally form into two pseudo-Dirac pairs of 
fields, leaving only one almost pure Majorana active neutrino 
and one conventionally very heavy sterile Majorana neutrino. We 
further find that the naturally strong mixing between the active 
and sterile parts of the two pseudo-Dirac pairs can easily affect 
the mixing between active neutrinos which is otherwise small, 
i.e., that due to the Dirac mass matrix induced mixing 
analogous to what is known to occur in the quark sector,  
amplifying such small mixing between active neutrinos to much 
larger values. 

We have chosen a limited relative value of the sterile neutrino 
mass scale, $M$, that allows for easy discernment of the nature 
of the effects. It should be noted, however, that the primary 
effect of increasing $M$ is to decrease the separation of the 
two members of each pseudo-Dirac pair while producing the usual 
see-saw behavior for the remaining pair. Thus, tiny differences 
in mass between masses that are not especially small themselves 
are, in the usual sense of the term, natural in this approach. 
This is contrary to the general expectation that the small mass 
differences responsible for the observed neutrino oscillation 
phenomena presage small absolute masses for all of the neutrinos. 

The above is most easily discerned in the case when the Dirac 
mass terms for the neutrinos are well separated in value. It 
remains conceivable that, if their differences are small for 
some other reason, then the splitting between the pseudo-Dirac 
pairs may be larger than that between flavors. In this case, 
it is still true that large flavor mixing is naturally induced.

\section{Acknowledgments}
This research is supported in part by the Department of Energy under
contract W-7405-ENG-36, in part by the National Science Foundation 
under NSF Grant \# PHY0099385 and in part by the Australian Research 
Council.

\newpage

\noindent TABLE: Eigenmasses for various values of $\theta$ and $\phi$
for cases of two approximately degenerate pseudo-Dirac pairs.
\begin{verbatim}
__________________________________________________________________________
\end{verbatim}
$\theta =  9.324078$,   $\phi =  0$

\begin{verbatim}
      mass    1active  2active   3active   1sterile  2sterile  3sterile

   1.099328   0.635032   0.000000 -0.310533  0.698108  0.000000 -0.113793
   1.100680  -0.633620   0.000000  0.314383  0.697413  0.000000 -0.115345
   1.100000   0.000000   0.707107  0.000000  0.000000  0.707107  0.000000
   1.100000   0.000000  -0.707107  0.000000  0.000000  0.707107  0.000000
   0.007438   0.441883   0.000000  0.897064 -0.003287  0.000000 -0.002224
1000.008789   0.000162   0.000000  0.002960  0.162017  0.000000  0.986784
\end{verbatim}

$\theta =  9.324078$,   $\phi =  2.25$

\begin{verbatim}
    mass      1active   2active   3active  1sterile  2sterile  3sterile

   1.095953   0.479130  -0.468214  -0.225940  0.525106 -0.466489 -0.082539
   1.096608   0.437964  -0.514829  -0.208027 -0.480274  0.513243  0.076041
   1.103359   0.416946   0.529767  -0.212981  0.460041  0.531383 -0.078333
   1.104056  -0.458049  -0.484588   0.235669  0.505710  0.486376 -0.086730
   0.007438   0.441553   0.015769   0.897088 -0.003285 -0.000109 -0.002224
1000.008789   0.000162   0.000007   0.002960  0.161892  0.006361  0.986784
\end{verbatim}

$\theta =  9.324078$,   $\phi =  4.5$

\begin{verbatim}
    mass      1active   2active   3active  1sterile  2sterile  3sterile

   1.092254  0.479875 -0.471453 -0.217491  0.524155 -0.468127 -0.079183
   1.092888 -0.458763  0.495815  0.209390  0.501371 -0.492614 -0.076279
   1.107010  0.416602  0.526536 -0.221472  0.461189  0.529886 -0.081725
   1.107726  0.437718  0.503654 -0.234323 -0.484866 -0.507196  0.086521
   0.007439  0.440571  0.031517  0.897156 -0.003273 -0.000217 -0.002226
1000.008789  0.000162  0.000014  0.002960  0.161517  0.012712  0.986784
\end{verbatim}

$\theta =  9.324078$   $\phi =  22.5$

\begin{verbatim}
    mass      1active   2active   3active  1sterile  2sterile  3sterile

    1.062925  0.550356 -0.405921 -0.179528  0.584987  -0.392239 -0.063608
    1.063381  0.546548 -0.411257 -0.179609 -0.581185   0.397574  0.063663
    1.134871  0.337840  0.568726 -0.249457  0.383405   0.586755 -0.094367
    1.135731  0.341702  0.564710 -0.254038 -0.388074  -0.583058  0.096172
    0.007475  0.409265  0.154109  0.899298 -0.003058  -0.001048 -0.002241
1000.008789  0.000150   0.000068  0.002960  0.149684   0.062001  0.986784
\end{verbatim}

$\theta =  9.324078$   $\phi =  45$

\begin{verbatim}
     mass    1active   2active   3active   1sterile  2sterile  3sterile

   1.030458  0.632073 -0.290233 -0.127244  0.651329 -0.271878 -0.043708
   1.030692 -0.630801  0.292859  0.127989  0.650162 -0.274406 -0.043972
   1.163620  0.226485  0.612428 -0.270973  0.263544  0.647849 -0.105102
   1.164612  0.227955  0.610618 -0.274587 -0.265472 -0.646488  0.106595
   0.007566  0.315141  0.286490  0.904762 -0.002384 -0.001969 -0.002282
1000.008789  0.000115  0.000126  0.002960  0.114563  0.114563  0.986784
________________________________________________________________________
\end{verbatim}

\begin{figure*}[h] 
\includegraphics[height=4.0in]{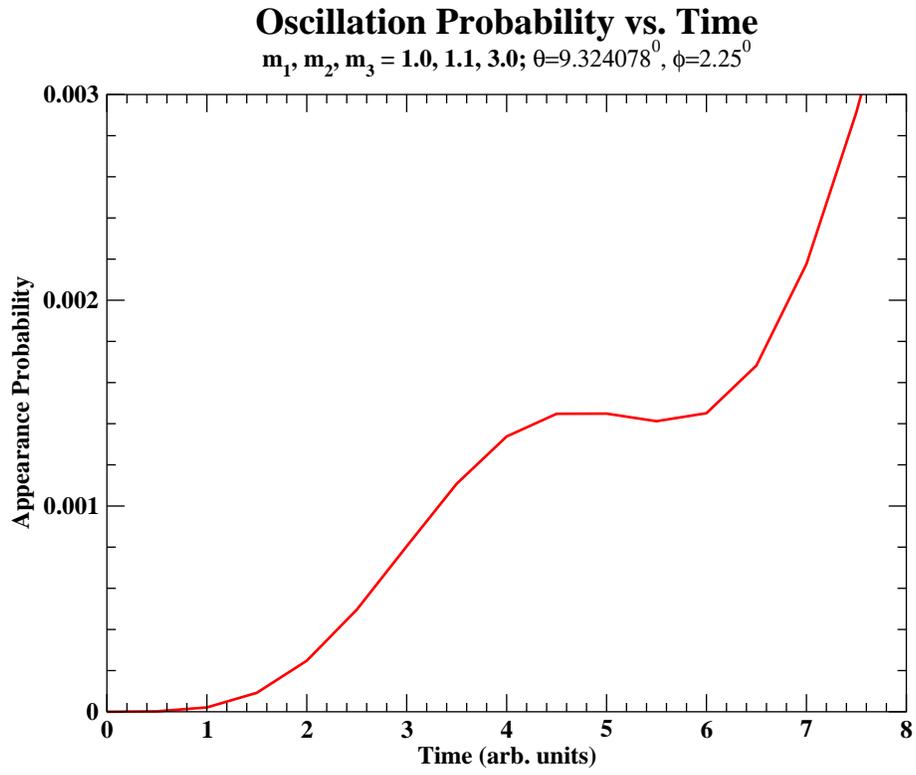} 
\caption{ Appearance probability for one neutrino 
flavor using the second set of angle parameters 
in the Table.} 
\label{appear} 
\end{figure*} 


\begin{thebibliography}{99}
\bibitem{fn1} {In much of the literature, the sterile space is 
	referred to as ``Right-handed'' (or just R) and the 
	active space as ``Left-handed'' (or just L), which 
	follows from the behavior of the components of a Dirac 
	neutrino where the neutrino is defined as that neutral 
	lepton emitted along with a positively charged lepton. 
	Since we are dealing with mass matrices, which necessarily 
	all couple left-handed representations to right-handed 
	representations, we choose to refer to the Weyl neutrino 
	representations as active and sterile. ``Sterile'' refers 
	only to the SM; these neutrino states may have non-SM 
	interactions.}
\end{thebibliography}
\end{document}